\documentclass[twocolumn,pra,aps,floatfix,superscriptaddress]{revtex4}
\usepackage{epsfig,graphicx}
\usepackage{flafter}
\usepackage{amsmath}
\usepackage{color}
\usepackage{bm}
\usepackage{amssymb}
\usepackage{epstopdf}
\usepackage{amsmath}
\usepackage{amstext}
\usepackage{amsthm}
\usepackage{amsfonts}
\usepackage{latexsym}
\usepackage{multirow}
\usepackage[utf8]{inputenc}
\usepackage{soul}

\begin{document}

\title{Mode-division (de)multiplexing using adiabatic passage and supersymmetric waveguides}

\author{G. Queralt\'{o}}
\affiliation{Departament de F\'{\i}sica, Universitat
Aut\`{o}noma de Barcelona, E--08193 Bellaterra, Spain}
\author{V. Ahufinger}
\affiliation{Departament de F\'{\i}sica, Universitat
Aut\`{o}noma de Barcelona, E--08193 Bellaterra, Spain}
\author{J. Mompart}
\affiliation{Departament de F\'{\i}sica, Universitat
Aut\`{o}noma de Barcelona, E--08193 Bellaterra, Spain}

\date{\today}

\begin{abstract}
The development of mode-division multiplexing techniques is an important step to increase the information processing capacity. In this context, we design an efficient and robust mode-division (de)multiplexing integrated device based on the combination of spatial adiabatic passage and supersymmetric techniques. It consists of two identical step-index external waveguides coupled to a supersymmetric central one with a specific modal content that prevents the transfer of the fundamental transverse electric spatial mode. The separation between waveguides is engineered along the propagation direction to optimize spatial adiabatic passage for the first excited transverse electric spatial mode of the step-index waveguides. Thus, by injecting a superposition of the two lowest spatial modes into the step-index left waveguide, the fundamental mode remains in the left waveguide while the first excited mode is fully transmitted to the right waveguide. Output fidelities $\mathcal{F}>0.90$ are obtained for a broad range of geometrical parameter values and light's wavelengths, reaching $\mathcal{F}=0.99$ for optimized values.
\end{abstract}

\maketitle

\section{Introduction}

\begin{figure*}[t!]
\centering
\includegraphics[width=0.9\linewidth]{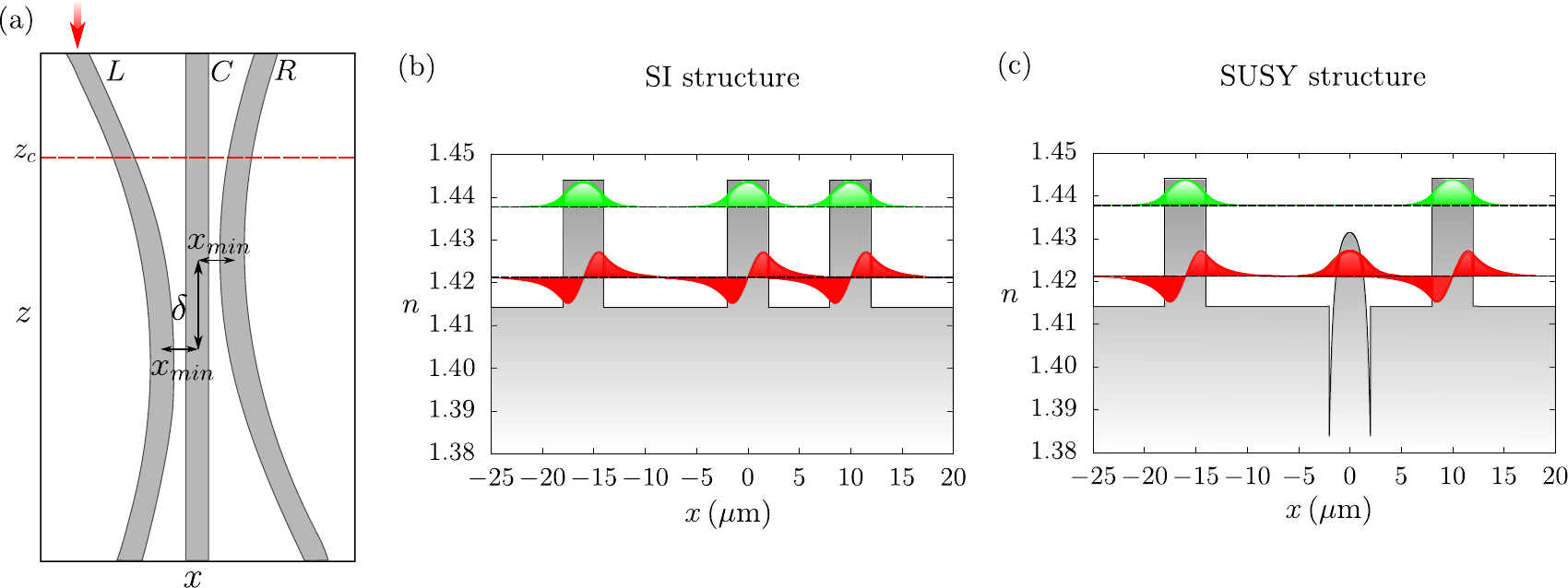}
\caption{(a) Schematic representation of the proposed device viewed from above. The width of the waveguide cores is $d=4 \, \mu \rm{m}$, the minimum separation between waveguides is $x_{min}= 7 \, \mu \rm{m}$, the radius of curvature of the external waveguides is $r=3.5 \, \rm{m}$, the spatial delay between their centers is $\delta=4 \, \rm{mm}$ and the total length along $z$ is $D=18\, \rm{mm}$. Refractive index distribution and transverse mode profiles of each waveguide at $z_c$ in (a) for (b) the SI structure and (c) the SUSY structure.}
\label{fig:device}
\end{figure*}

Integrated optical devices exhibiting both high fidelity and high speed transmission are expected to foster novel communication platforms \cite{Agrell2016} paving the way for scalable photonic quantum technologies \cite{Obrien2009,Meany2015}. Moreover, the emerging technology of Space-Division Multiplexing \cite{Li2014,Richardson2013} has recently attracted a lot of attention due to the increasing demand of high-capacity optical transmissions and the proximity of the capacity crunch, being Mode-Division Multiplexing (MDM) one of the proposed solutions to take profit of the spatial degrees of freedom working with multimode fibers and waveguides \cite{Richardson2016}. In this context, integrated MDM devices \cite{Dai2017} are being developed using many different approaches including photonic lanterns \cite{Birks2015}, multimode interference \cite{Uematsu2012}, asymmetric directional \cite{Ding2013,Luo2014} and adiabatic \cite{Xing2013,Zhang2017} couplers, asymmetric y-junctions \cite{Driscoll2013,Chen2016}, microring-based \cite{Dorin2014} and Supersymmetric (SUSY) optical devices \cite{Heinrich2014-1}. 

SUSY, discovered in the 70's and originally developed to unify the mathematical treatment of bosons and fermions, is presently applied to many areas of Physics \cite{Dine2007}. In particular, due to the analogies between the time-independent Schr\"odinger and Helmholtz equations, SUSY has recently been extended from non-relativistic Quantum Mechanics \cite{Cooper1995} to Helmholtz Optics \cite{Chumakov1994}. In fact, SUSY techniques offer, through an appropriate manipulation in space of the refractive index, new ways to control the modal content of light beams in optical waveguides \cite{Miri2013-2,Principe2015}, to design refractive index landscapes with non-trivial properties \cite{Miri2013-1,Miri2014} or to engineer different index profiles with identical scattering characteristics \cite{Heinrich2014-2,Longhi2015}. Regarding MDM, SUSY-based optical devices constitute a promising alternative to standard MDM devices since they offer global phase-matching and efficient mode conversion \cite{Heinrich2014-1} in an integrated and scalable way, with low losses and compatible with other existing multiplexing techniques. However, the main drawback of most SUSY optical devices is their lack of robustness against parameter values variations and their sensitivity to experimental imperfections. 

On the other hand, Spatial Adiabatic Passage (SAP) techniques \cite{Menchon2016} have been proposed and experimentally reported as a high-efficient and robust method to transfer a light beam between the outermost waveguides in a system of three identical evanescently-coupled waveguides \cite{Longhi2007,Menchon2012}, providing a new set of robust photonic devices \cite{Dreisow2009,Ciret2013,Menchon2013,Ng2017}.

In this article, we propose to combine SUSY and SAP techniques to design an efficient and robust device which can be used for multiplexing/demultiplexing spatial modes, to manipulate and study the modal content of an input field distribution or to filter signals and remove non-desired modes. Here, we demonstrate that a system of three coupled waveguides, with two identical step-index external waveguides and a supersymmetric central one, engineered along the propagation direction to optimize SAP for the first excited spatial mode of the step-index waveguides, can be used to demultiplex a superposition of the two lowest $(m=0,1)$ transverse electric $(TE_m)$ spatial modes. Note that, using the output ports as the input ones, the same device can operate as a multiplexer. Although the idea of SUSY has already been experimentally applied for MDM purposes \cite{Heinrich2014-1}, its combination with SAP techniques is proposed for the first time to our knowledge leading to a great improvement in terms of robustness and efficiency.  

\section{Physical system}

We consider a set of three planar evanescently-coupled waveguides consisting of a core refractive index $n_{core}$ embedded in a medium of lower refractive index $n_{clad}$. The central ($C$) waveguide is straight while the left ($L$) and right ($R$) waveguides are truncated circles of radius $r$ with their centers displaced $\delta$ from each other along the $z$ direction, see Fig. \ref{fig:device}(a). Regarding the refractive index profiles, two structures will be investigated: (i) the step-index (SI) structure formed by three identical step-index waveguides with refractive index $n_j(x)$ where $j=L,C,R$, see Fig. \ref{fig:device}(b), and (ii) the SUSY structure consisting of two identical step-index waveguides with refractive index $n_L(x)=n_R(x)$ and a superpartner central one characterized by $n^{(p)}_C(x)$, see Fig. \ref{fig:device}(c).

To be specific, we consider $n_{core}=1.444$ and $n_{clad}=1.414$, corresponding to refractive indices available on fused silica at telecom wavelength $\lambda=1.55 \, \mu \rm{m}$ which is the material being used in state-of-the-art SUSY waveguides \cite{Heinrich2014-1}. However, the proposal described here is not restricted to specific refractive index contrasts and profiles \cite{Miri2014-3,Miri-tesis}. The width $d$ of the step-index waveguides is chosen to allow the propagation of the fundamental $TE_0$ and the first excited $TE_1$ modes while the superpartner waveguide only supports one mode with the same propagation constant as the $TE_1$ mode of the step-index waveguides, see Fig. \ref{fig:device}(c). The propagation of the TE component of the electric field for each waveguide $j$ can be described by the Helmholtz equation:
\begin{equation}
\left[\frac{\partial^2}{\partial x^2}+\frac{\partial^2}{\partial z^2}+k_0^2 n_j^2 (x) \right]E^j_y(x,z)=0,
\label{eq:Helmhjoltz}
\end{equation}
where $k_0$ is the vacuum wavenumber. The electric field in each waveguide can be expressed as a superposition of modes: 
\begin{equation}
E^j_y(x,z)=\sum_{m} a^j_m(z)f^j_m(x) \exp[i\beta^j_m z],
\label{eq:Efield}
\end{equation}
where $a^j_m(z)$ is the amplitude, $f^j_m(x)$ is the transverse spatial distribution and $\beta^j_m$ the propagation constant of mode $m$. 

\section{Theoretical background}

Light propagation in a system of three weakly coupled identical waveguides with $\beta_m^L=\beta_m^C=\beta_m^R$ can be described through coupled-mode equations \cite{Chen2007}. For each mode $m$ these equations read:
\begin{equation}
i\frac{\rm{d}}{\rm{d}z} \left(
\begin{array}{l}
a_m^{L} \\
a_m^{C} \\ 
a_m^{R}
\end{array}
\right)= \frac{1}{2}\left(
\begin{array}{ccc}
2\beta_m^L & -\Omega^{LC}_{m} & 0 \\
-\Omega^{LC}_{m} & 2\beta_m^C & -\Omega^{RC}_{m}\\ 
0 & -\Omega^{RC}_{m} & 2\beta_m^R
\end{array}
\right)\left(
\begin{array}{l}
a_m^{L} \\
a_m^{C} \\ 
a_m^{R}
\end{array}
\right),
\label{eq:light}
\end{equation}
where we have assumed that left and right waveguides are not being directly coupled and $\Omega_m^{LC}(z)$ $(\Omega_m^{RC}(z))$ is the coupling coefficient between mode $m$ of the left (right) and central waveguides. For the SI structure, we have two independent sets of equations, one for each mode $(m=0,1)$ while for the SUSY structure there is only coupling between the $TE_1$ mode of the step-index external waveguides and the fundamental mode of the superpartner central one. Note that, since $\beta_1^{L}=\beta_0^{(p)C}=\beta_1^{R}$ for the SUSY structure, the corresponding modes can be perfectly phase-matched \cite{Heinrich2014-1}. 

\subsection{Spatial adiabatic passage techniques} 

Diagonalizing the matrix of the right hand side of Eq. \ref{eq:light}, one obtains that one of the eigenvectors (supermodes) of the system only involves light in the left and right waveguides:
\begin{equation}
D(\Theta_m)= \left(
\begin{array}{c}
\cos(\Theta_m) \\
0 \\ 
-\sin(\Theta_m)
\end{array}
\right),
\label{eq:supermode}
\end{equation}
where $\Theta_m$ is the mixing angle given by $\tan \Theta_m \equiv \Omega_m^{LC}/\Omega_m^{RC}$. 
This supermode is analogous to the well known dark state of quantum optics \cite{Bergmann1998,Vitanov2017}.  If the input beam coincides with the dark supermode, its adiabatic modification allows for an efficient and robust transfer of the light beam between the outermost waveguides without light propagation into the central one, this process is known as SAP of light \cite{Longhi2007}. In particular, we consider that light is injected into the left waveguide and the couplings are engineered following a counter-intuitive sequence along $z$. By counter-intuitive sequence we mean that, first, the right waveguide approaches to the central waveguide and then, with a certain spatial delay, the left waveguide approaches to the central waveguide whereas the right separates from it, see Fig. \ref{fig:device}(a). With this spatial configuration, the mixing angle $\Theta_m$ evolves from $0$ to $\pi/2$ and light is transferred from the left to the right waveguide. The variation of the couplings along $z$ is engineered to follow a Gaussian function of the form \cite{Menchon2012}:
\begin{equation}
\Omega_m^{LC,RC}(z)\approx \Omega_m(x_{min},\lambda)\exp\left[\frac{-(z-D/2\pm \delta/2)^2}{2rl_m(\lambda)}\right],
\label{eq:omega3}
\end{equation}
where $l_{m}(\lambda)$ is a decaying constant and $\Omega_m(x_{min},\lambda)\equiv \widetilde{\Omega}_{m}(\lambda)\exp[-x_{min}/l_m(\lambda)]$ gives the maximum value of the couplings. On the other hand, the global adiabaticity condition reads:
\begin{equation}
\delta\sqrt{(\Omega_m^{LC})^2+(\Omega_m^{RC})^2} > A,
\label{eq:adiabaticity}
\end{equation}
where $A$ is a dimensionless constant that takes a value around 10 for optimal parameter values such as the spatial delay
between the couplings \cite{Bergmann1998}. Thus, Eqs. \ref{eq:omega3} and \ref{eq:adiabaticity} allow one to select the appropriate geometrical parameter values $\{D,\delta,x_{min}\}$ to efficiently perform SAP of mode $m$. 

\begin{figure}[h!]
\centering
\includegraphics[width=0.75\linewidth]{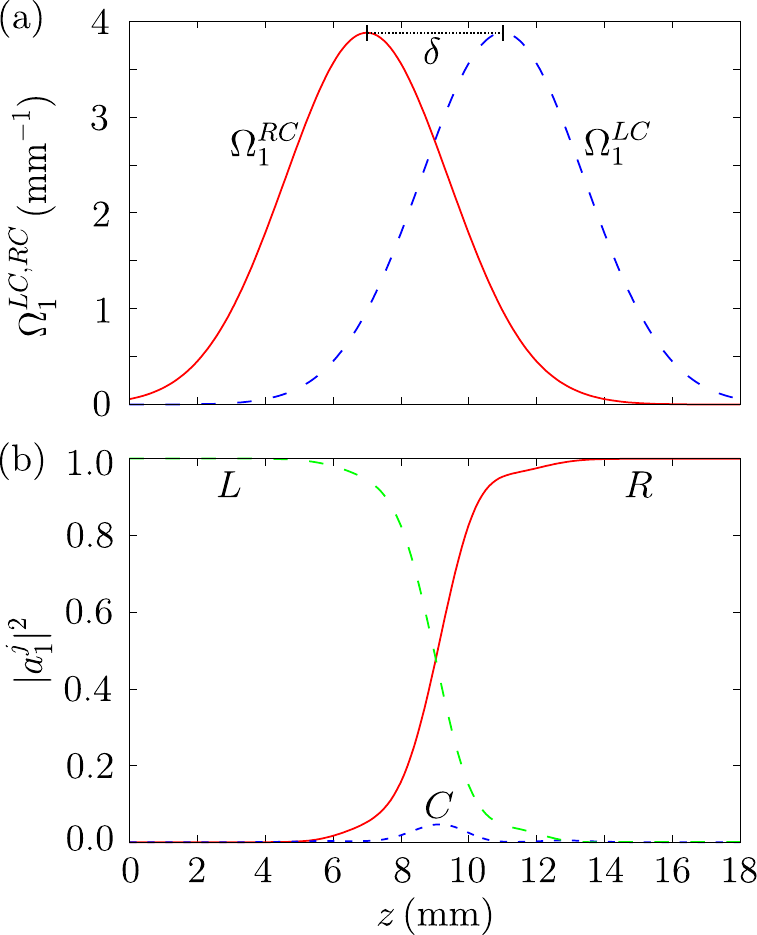}
\caption{(a) Spatial sequence of the coupling coefficients for $m=1$ in the considered device of Fig. \ref{fig:device}(a) with the SI structure. (b) Evolution of the $TE_1$ mode intensity along $z$ in each waveguide of the SI structure as predicted by the coupled-mode equations. Parameter values: $\lambda=1.55 \, \mu \rm{m}$, $r= 3.5 \, \rm{m}$, $\delta=4 \, \rm{mm}$, $x_{min}=7 \, \mu \rm{m}$, $D=18 \, \rm{mm}$, $\widetilde{\Omega}_1=263 \, \rm{mm^{-1}}$ and $l_1=1.66 \, \mu \rm{m}$.}
\label{fig:pulses}
\end{figure}

\subsection{Supersymmetric waveguides} 

SUSY techniques allow to systematically construct a superpartner profile that shares all the propagation constants with the original profile except for the fundamental mode, obtaining global phase matching for an arbitrarily large number of modes [21]. In our case, to design the superpartner index profile $n^{(p)}_C(x)$ for the central waveguide that  only propagates one mode with $\beta_0^{(p)C}$ equal to $\beta^C_1$ of the SI structure, we perform a systematic deformation of $n_C(x)$ following SUSY techniques (see \cite{Miri2013-2,Miri2014-3} for more details). From Eq. \ref{eq:Helmhjoltz}, one can easily derive the eigenvalue equation:
\begin{equation}
\mathcal{H}_Cf^C_m(x)=-(\beta^C_m)^2 f^C_m(x),
\label{eq:eigenvalue}
\end{equation}
where $\mathcal{H}_C=-\rm{d}^2/\rm{d} x^2-k_0^2n^2_C(x)$. The analogy with the time independent Schr\"odinger equation $\mathcal{H}\Psi_m (x)=\varepsilon_m \Psi_m (x)$ allows the application of SUSY techniques to Helmholtz optics \cite{Chumakov1994}. The superpartner profile can be derived factorizing the Hamiltonian in terms of
$\mathcal{H}_C=A^{\dag}A-(\beta_0^C)^2$ and $\mathcal{H}_C^{(p)}=AA^{\dag}-(\beta_0^C)^2$, where $A=d/dx+W(x)$ and $A^{\dag}=-d/dx+W(x)$ and $W(x)$ is the so-called superpotential, associated with the mode that needs to be removed:
\begin{equation}
W(x)=-\frac{d}{dx}\ln f^C_0(x).
\label{eq:superpotential}
\end{equation}
Using these relations, one finds that the superpartner index profile satisfies:
\begin{equation}
n_C^{(p)}(x)=\frac{1}{k_0}\sqrt{(\beta_0^C)^2-W(x)^2 - \frac{\rm{d}W(x)}{\rm{d}x}}.
\label{eq:index-of-refraction}
\end{equation}
Knowing the analytical expression of $f^C_0(x)$ for the step-index profile \cite{Chen2007}, the superpotential can be derived from Eq. \ref{eq:superpotential}. Substituting the result into Eq. \ref{eq:index-of-refraction}, one obtains the superpartner index profile \cite{Miri2014-3,Miri-tesis}:
\begin{equation}
n_C^{(p)}(x) = \left\lbrace 
\begin{array}{ll}
\sqrt{n_{core}^2-2\left(\frac{k_x}{k_0}\right)^2\sec^2(k_xx)}, & |x| \leq d/2 \\
n_{clad}, & |x| > d/2
\end{array},
\right.
\end{equation}
where $k_x$ is the wavevector component in the $x$ direction.

\section{Results and discussion}

\begin{figure*}[t!]
\centering
\includegraphics[width=0.9\linewidth]{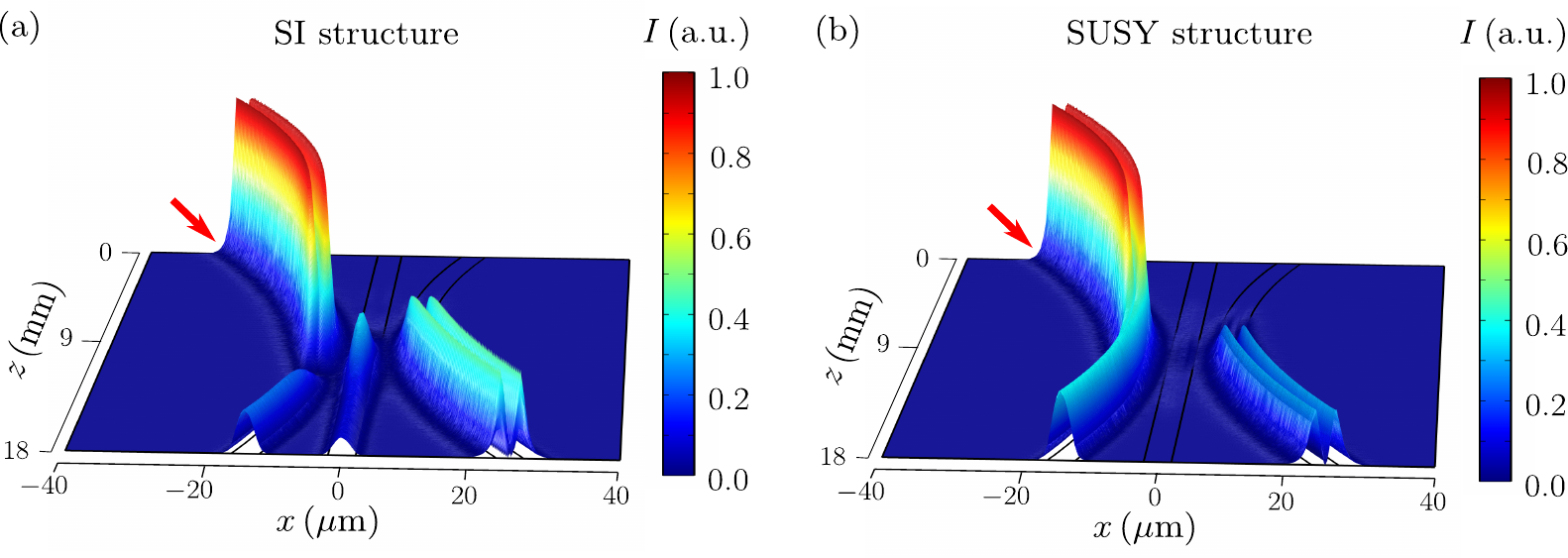}
\caption{Numerical simulations of light intensity propagation along $z$ when one injects an equally weighted superposition of the fundamental and the first excited modes in the left waveguide of (a) the SI structure and (b) the SUSY structure of Fig. \ref{fig:device}. Parameter values: $\lambda=1.55 \, \mu \rm{m}$, $d=4 \, \mu \rm{m}$,  $r= 3.5 \, \rm{m}$, $\delta=4 \, \rm{mm}$, $x_{min}=7 \, \mu \rm{m}$, and $D=18 \, \rm{mm}$.}
\label{fig:intensity}
\end{figure*}

In this section we will demonstrate that the combination of SAP and SUSY techniques allows for a very efficient and robust demultiplexing of transverse spatial modes. We will show that, by injecting an input beam consisting of an equally weighted superposition of $TE_0$ and $TE_1$ modes in the left waveguide, the $TE_1$ mode will be efficiently transferred to the right waveguide while the $TE_0$ mode will remain in the left one.

To guarantee an efficient and robust transfer of the $TE_1$ mode, we optimize the geometrical parameters to perform SAP of this mode in the SI structure. Note that, since the coupling for the $TE_0$ mode is weaker than for the $TE_1$ mode, the adiabaticty condition may not be fulfilled for this mode (see Eq. \ref{eq:adiabaticity}), preventing its efficient transfer. To characterize the dependence of the couplings along $z$, we numerically simulate the propagation of the $TE_1$ mode between two straight step-index waveguides separated different distances obtaining $\widetilde{\Omega}_1=263 \, \rm{mm^{-1}}$ and $l_1=1.66 \, \mu \rm{m}$, see discussion after Eq. \ref{eq:omega3}. Casting these values into Eq. \ref{eq:omega3} allows to find the optimized set of parameters to efficiently perform SAP of $TE_1$, see Fig. \ref{fig:pulses}(a). Integrating Eq. \ref{eq:light} for these optimized parameters, one obtains that the $TE_1$ mode is efficiently transferred to the right waveguide for the SI structure, see Fig. \ref{fig:pulses}(b). As $\beta_0^{(p)C}=\beta_1^L=\beta_1^R$, the optimized geometry will also provide maximum $TE_1$ mode transmission for the SUSY structure.

Performing full numerical simulations using Finite Difference Methods for the SI structure with the optimized geometry, we confirm the results of the coupled-mode equations, obtaining that $99.3\%$ of the injected intensity of the $TE_1$ mode is transmitted to the right waveguide. For the $TE_0$ mode, the injected intensity is spread among the three waveguides obtaining $11.2\%$ in the right, $25.9\%$ in the central and $62.8\%$ in the left waveguides, as shown in Fig. \ref{fig:intensity}(a). To keep the $TE_0$ mode in the left waveguide and achieve efficient mode separation, we now consider the SUSY structure. The numerical simulations for the SUSY structure with the optimized geometry confirm that $99.3\%$ of the injected intensity of the $TE_1$ mode  is transmitted to the right waveguide while $99.8\%$ of the injected intensity of the $TE_0$ mode remains in the left waveguide, see Fig. \ref{fig:intensity}(b). To investigate the efficiency and robustness of the proposed device for demultiplexing purposes we introduce a Figure of Merit of the form:
\begin{equation}
\mathcal{F}=\frac{I^L_{0,out}}{I^L_{0,in}} \cdot \frac{I^R_{1,out}}{I^L_{1,in}}=\widetilde{I}^L_0 \cdot \widetilde{I}^R_1,
\label{eq:Figure-of-merit}
\end{equation}
where $\widetilde{I}^L_0$ is the fraction of intensity of the $TE_0$ mode that remains in the left waveguide and $\widetilde{I}^R_1$ is the fraction of intensity of the $TE_1$ mode that is transmitted to the right waveguide. From now on, we will consider that we have efficient demultiplexing for fidelity values fulfilling $\mathcal{F}>0.90$. The maximum fidelity $\mathcal{F}=0.99$ is achieved for the parameter values used so far to optimize SAP for the $TE_1$ mode. 

\begin{figure*}[t!]
\centering
\includegraphics[width=0.9\linewidth]{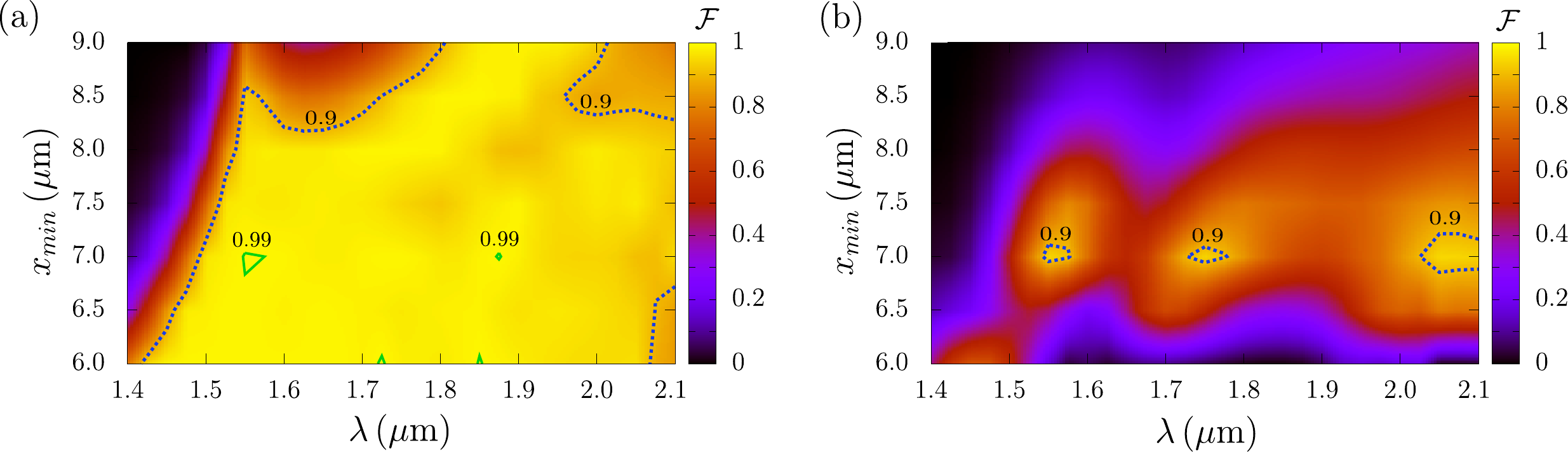}
\caption{Demultiplexing fidelity calculated through numerical simulations when we inject an equally weighted superposition of the fundamental and the first excited modes in the left waveguide for different wavelengths and minimum separation between waveguides for (a) the SUSY structure of Fig. \ref{fig:device} and (b) the straight SUSY structure of total length $\widetilde{D} = 2 \, \rm{mm}$.}
\label{fig:robustness}
\end{figure*}

To prove the robustness of the device, we perform numerical simulations varying the geometrical parameters and light's wavelength. From these simulations, we find that the device can be miniaturized down to $D_{min}=10 \, \rm{mm}$ and the spatial delay engineered within $3.5 \, \rm{mm} \leq$ $\delta$ $\leq 4.5 \, \rm{mm}$ maintaining a fidelity $\mathcal{F}=0.99$. Moreover, high fidelities $\mathcal{F}>0.90$ are achieved in a huge range $2.0 \, \rm{mm} < \delta < 6.0 \, \rm{mm}$. Outside this range, the fidelity decreases fast because once SAP conditions are not fulfilled, light is not efficiently transmitted from the left to the right waveguide and spreads between waveguides \cite{Longhi2007}. Figure \ref{fig:robustness}(a) shows the fidelity of the SUSY structure as a function of the light's wavelength $\lambda$ and the minimum separation between waveguides $x_{min}$ considering the refractive index profiles fixed and independent of $\lambda$. Efficient demultiplexing $(\mathcal{F}>0.90)$ is achieved in a broad region of $\Delta \lambda \sim 0.5 \, \mu \rm{m}$ and $\Delta x_{min} \sim 2 \, \mu \rm{m}$, reaching $\mathcal{F}=0.99$ in some small regions. Although we have considered fixed refractive indices, the influence of dispersion remains negligible across $1.5\, \rm{\mu m} \leq \lambda \leq 1.6 \, \rm{\mu m}$ \cite{Heinrich2014-1} and we have numerically checked that it introduces a deviation of the fidelities below $3\%$ for the range $1.6 \, \rm{\mu m} < \lambda < 2 \, \rm{\mu m}$. For longer wavelengths, the impact of dispersion becomes important. Losses due to the bending of the waveguides associated to the adiabatic passage geometry  \cite{Menchon2013} and the ones associated to SUSY mode conversion \cite{Heinrich2014-1} are negligible and we only have the usual propagation losses which depend on the characteristics of the specific waveguide.

To highlight the advantages of combining SUSY and SAP techniques, we compare the designed device with a device consisting of three straight waveguides with two external identical step-index waveguides with refractive index $n_L(x)=n_R(x)$ and a superpartner central one with $n^{(p)}_C(x)$. The straight SUSY structure allows demultiplexing with smaller devices $(\tilde{D} \sim 2 \, \rm{mm})$ than the here investigated device, since it does not require adiabaticity. Nevertheless, its efficiency rapidly decreases in front of small parameter variations, see Fig. \ref{fig:robustness}(b). With the straight configuration, the same level of fidelity as for the SAP geometry cannot be achieved and the regions where one has efficient demultiplexing $(\mathcal{F}>0.90)$  are drastically reduced to $\Delta \lambda < 0.05 \, \mu \rm{m}$ and $\Delta x_{min} < 0.5 \, \mu \rm{m}$.

Although we have designed the device to operate at telecom wavelengths due to the high-fidelity mode conversion that SUSY offers over the telecommunication C-band \cite{Heinrich2014-1}, it can be optimized to work at different wavelengths by simply modifying the geometrical parameter values. Moreover, as the proposed device is efficient for a broad wavelength range, it may also be used for demultiplexing of light pulses and it is fully compatible with wavelength division multiplexing. In addition, despite our investigations were based on step-index waveguides to work with analytical expressions for the superpartner profile, SUSY techniques could be applied to other index profiles \cite{Miri2014-3,Miri-tesis} leading to smoother superpartners in which the sharp refractive index dip is reduced and could be more easily experimentally implemented \cite{Heinrich2014-1}. 

As a proof of principle, we have focused on the simplest possible case for which only two $TE$ spatial modes can propagate through the step-index planar waveguides. However, this configuration can be generalized to a higher number of TE modes as SUSY techniques offer global phase matching, to transverse magnetic modes or even to orbital angular momentum modes in optical fibers \cite{Miri2013-2}. Furthermore, if we design a step-index waveguide that supports $N$ modes, each guided mode will have a counterpart in the superpartner central waveguide with exactly the same propagation constant except for the fundamental one. In this case, the index profile of the central waveguide can be engineered by applying reiteratively SUSY transformations to select the modes to be transferred between the outermost waveguides. Finally, more complex devices can be constructed to demultiplex $N$ spatial modes by coupling in series different devices in an efficient and robust way, similarly to the hierarchical ladders of supersymmetric structures \cite{Heinrich2014-1}. Note that, although demultiplexing between two modes could also be achieved using a central step-index waveguide designed in such a way that only one mode propagates with the same propagation constant as the first excited mode of the external waveguides, the extension to higher order modes cannot be performed.

\section{Conclusions}

We have demonstrated that it is possible to design an efficient and robust demultiplexing device by combining supersymmetric structures and spatial adiabatic passage techniques. We have considered a triple-waveguide system with two curved step-index waveguides with their centers displaced from each other and a central straight waveguide characterized by a superpartner index profile. SUSY techniques have allowed to design the superpartner profile with the desired modal content and SAP techniques have provided robustness to the device. 

In particular, we have demonstrated that by injecting an input beam consisting of an equally weighted superposition of the $TE_0$ and $TE_1$  modes in the left waveguide, the $TE_1$ mode is transferred to the right waveguide while the $TE_0$ mode remains in the left one. Moreover, we have also tested the efficiency and robustness of the device obtaining demultiplexing fidelities $\mathcal{F}>0.90$ in a broad region of $\Delta \lambda \sim 0.5 \, \mu \rm{m}$ and $\Delta x_{min} \sim 2 \, \mu \rm{m}$, reaching $\mathcal{F}=0.99$ for optimized values. 

The proposed device offers a significant improvement in terms of robustness compared to previous approaches using SUSY for mode filtering \cite{Miri2013-2,Heinrich2014-1} and confirms that supersymmetric structures can be used for multiplexing/demultiplexing spatial modes, to manipulate and study the modal content of an input field distribution or to filter signals and remove the non-desired modes in an efficient and integrated way. In addition, the high obtained fidelities open promising perspectives in the field of quantum integrated photonics to, for instance, prepare and manipulate quantum states with minimal errors or by taking profit of the high dimensional Hilbert space associated to spatial modes \cite{Mohanty2017}.  


\section{AKNOWLEDGMENTS}

The authors gratefully acknowledge financial support through the Spanish Ministry of Science and Innovation (MINECO) (Contract No. FIS2014-57460P) and the Catalan Government (Contract No. SGR2014-1639).

\end{document}